\documentclass[showpacs,preprintnumbers,amsmath,amssymb,twocolumn,showkeys]{revtex4}
\usepackage{dcolumn}    
\usepackage{bm}         
\usepackage{ifpdf}
\usepackage{graphicx,amssymb,lineno}
\usepackage{amssymb,lineno,amsfonts}
\usepackage{graphicx}   
\usepackage{dcolumn}    
\usepackage{bm}         
\usepackage{bbm}
\usepackage{mathrsfs}
\usepackage{upgreek}
\usepackage{mathtools}

\begin{document}

\title{NMR implementation of Quantum Delayed-Choice Experiment}

\author{
Soumya Singha Roy, Abhishek Shukla, and T. S. Mahesh
}
\email{mahesh.ts@iiserpune.ac.in}
\affiliation{
Department of Physics and NMR Research Center,Indian Institute of Science Education and Research, Pune 411008, India 
}

\date{\today}

\begin{abstract}
{
We report the first experimental demonstration of quantum delayed-choice experiment via
nuclear magnetic resonance techniques.  
Two spin-1/2 nuclei from each molecule of a liquid ensemble are used as target and ancilla
qubits.  The circuit corresponding to the recently proposed quantum delayed-choice setup 
has been implemented with different states of ancilla qubit.  As expected in theory, our experiments 
clearly demonstrate continuous morphing of the target qubit between particle-like and wave-like behaviors.  
The experimental visibility of the interference patterns shows good agreement with the theory.
}
\end{abstract}

\maketitle

\section{Introduction}
``Is light made up of waves or particles?" has been an intriguing question 
over past many centuries, and the answer remains a mystery even today.
The first comprehensive wave theory of light was advanced by Huygens
\cite{Huygens}. 
He demonstrated how waves might interfere to form a wavefront propagating 
in a straight line, and he could also explain reflection and refraction of
light.  Soon Newton could explain these
properties of light using corpuscular theory, in which light 
was made up of discrete particles \cite{newton}.  The corpuscular theory held over a
century till the much celebrated Young's double slit experiment clearly established
the wave theory of light \cite{young}.
In the Young's experiment a monochromatic beam of light passing through an obstacle
with two closely separated narrow slits produced an interference pattern
with troughs and crests just like one would expect if waves from two
different sources would interfere.  Other properties of light like 
diffraction and polarization could also be explained easily using the
wave theory. 
The 20th century developments such as Plank's theory of 
black-body radiation and Einstein's theory of photoelectric effects required 
quantization of light into photons \cite{planck,einsteinphoto}.  But the question remained whether
individual photons are waves or particles. Subsequent development of 
quantum mechanics was based on the notion of wave-particle duality \cite{greiner}, which was
essential to explain the behavior not only of the light quanta, but also 
of atomic and sub-atomic entities \cite{bohr}.  

\begin{center}
\begin{figure}
\includegraphics[width=7.5cm]{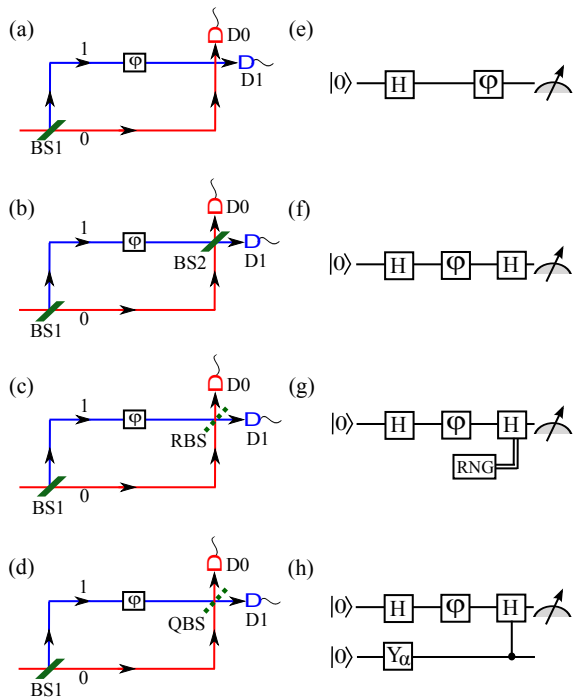}
\caption{Different types of Mach-Zehnder interferometer setups (a-d)
and equivalent quantum circuits (e-h). BS1 and BS2 are beam splitters,
$\phi$ is phase shifter, D0 and D1 are detectors.  RBS is a beam-splitter
switched ON or OFF by a random number generator (RNG) and  QBS is a beam-splitter
which is controlled by a quantum system in superposition.  In the quantum
circuits, $H$ is the Hadamard gate and $Y_\alpha = e^{-i \alpha \sigma_y}$
is used to prepare the state of ancilla qubit.
}
\label{fig1}
\end{figure}
\end{center}

The wave-particle duality of quantum systems is nicely illustrated
by a Mach-Zehnder interferometer (MZI) (see Fig. \ref{fig1})
\cite{zehnder,mach}.  The intensity
of the incident light is kept sufficiently week so that photons enter
the interferometer one by one.
In the open-setup (Fig. \ref{fig1}a), it consists
of a beam-splitter BS1, providing each incoming photon with two possible
paths, named 0 and 1.  A phase-shifter in path-1 introduces a relative
phase $\phi$ between the two paths.  The two detectors D0 and D1 help to 
identify the path traveled by the incident photon. Experimental 
results show that only one of the detectors
clicks at a time  \cite{aspect}.  Each click can then be correlated
with one of the two possible paths by attributing particle nature to the
photons.
Here the phase-shifter has no effect on the intensity of
the photons measured by either detector, and therefore no 
interference is observed in this setup.

In the closed-setup (Fig. \ref{fig1}b), 
the interferometer consists of a second beam-splitter BS2, which 
allows the two paths to meet before the detection.  Experimental results again show
that only one detector clicks at a time. But much to the astonishment
of  common intuition, the results after many clicks do show  
an interference pattern, i.e., the intensities recorded by each 
detector oscillates with $\phi$ \cite{aspect}.  Since only one photon is present
inside the interferometer at a time, each photon must
have taken both paths in the interferometer and therefore this setup 
clearly establishes the wave property of photons.

The naive question by the classical mindset is ``whether the photon
entering the interferometer decides to take one of the paths or 
both the paths depending on the experimental setup?".  
Scientists who believed 
in a deterministic nature had proposed that, unknown to the 
current experimentalist, there exists some extra information about state 
of the quantum system, which in principle dictates whether the
photon should take either path, or both the paths \cite{bohm}.  In other words,
they assumed some hidden information availed by the photon coming
out of BS1 about the existence or non-existence of BS2.  

In order to break this causal link between the two beam-splitters,
Wheeler proposed a modification in the MZI setup (Fig. \ref{fig1}c), 
in which the decision to introduce 
or not to introduce BS2 is to be made after the photon has already
passed through BS1 \cite{wheeler1,wheeler2,leggett}.  
This way, there is no causal connection 
between the selection of the paths by the photon and the presence of 
BS2.  Although initially considered as a `thought-experiment', this
proposal has recently been demonstrated by Jacques et al \cite{jacques}.  
In their experimental setup, the second beam-splitter (RBS) was controlled
by a random number generator (RNG), that choose to switch the beam-splitter
ON or OFF after the photon has already passed through BS1.
The results of this delayed-choice experiment was in agreement with 
Bohr's complementarity principle \cite{bohr}.  That is, the behavior of
the photon in the interferometer depends on the choice of the observable that is
measured, even when that choice is made at a
position and a time such that it is separated from
the entrance of the photon into the interferometer
by a space-like interval.
Breaking the causal link had no effect on
the results of the wave-particle duality, thus ruling out the existence
of hidden information \cite{jacques}. 

More recently, Ionicioiu and Terno have proposed a modified version
(Fig. \ref{fig1}d) of the
Wheeler's experiment which not only demonstrates the intrinsic duality, but
also shows that a photon can have a morphing behavior between particle and 
wave \cite{ionicioiu}. In their setup, BS2 is replaced with a beam splitter which is
switched OFF or ON depending on $\vert 0 \rangle$ or $\vert 1 \rangle$ state 
of a two-level quantum system.
Using this modification, Ionicioiu and Terno have been able to discard hidden 
variable theories which attempt to assign intrinsic wave or particle nature to individual 
photons even before the final measurement.  This proposed experiment is named
as `Quantum Delayed-Choice Experiment' \cite{ionicioiu}.

In this article we report the first experimental demonstration of quantum
delayed-choice experiment.
Using nuclear magnetic resonance (NMR) techniques we study the behavior of a target
spin-1/2 nucleus going through a similar situation as that of a photon going through an
interferometer.  Another spin-1/2 nucleus acts as an ancilla controlling 
the second beam-splitter.
In section II we briefly explain the theory and in section III
we describe the experimental results.

\section{Theory}
In the following we shall use the terminology of quantum information. 
The two possible paths of the interferometer are assigned with the orthogonal
states $\vert 0 \rangle$ and $\vert 1 \rangle$ of a quantum bit.  
The equivalent quantum circuits for the different setups of MZI
are shown in Figs. \ref{fig1}(e-h).  In these circuits the Hadamard operator H has the function of the beam splitter BS1.
It transforms the initial state $\vert 0 \rangle$  to the superposition
$(\vert 0 \rangle + \vert 1 \rangle)/\sqrt{2}$ such that both the states are now
equally probable. 
The detection operators for the two detectors are $D_0 = \vert 0 \rangle \langle 0 \vert$ and
$D_1 = \vert 1 \rangle \langle 1 \vert$.

In the open setup (Fig. \ref{fig1}e), the state after the phase shift 
becomes, $\vert \psi_p \rangle = (\vert 0 \rangle + e^{i\phi} \vert 1 \rangle)/\sqrt{2}$.    
The intensities recorded by the two detectors are given by the 
expectation values,
\begin{eqnarray}
S_{\mathrm{p},0} & = & \langle \psi_\mathrm{p} \vert D_0 \vert \psi_\mathrm{p} \rangle = \frac{1}{2} \;\; \mathrm{and} \nonumber \\
S_{\mathrm{p},1} & = & \langle \psi_\mathrm{p} \vert D_1 \vert \psi_\mathrm{p} \rangle = \frac{1}{2},
\label{sp0}
\end{eqnarray}
independent of the phase introduced.  Therefore no interference can  be observed
and accordingly this setup demonstrates the particle nature of the quantum system.
The visibility of the interference
\begin{eqnarray}
\nu = \frac{\mathrm{max}(S) - \mathrm{min}(S)}{\mathrm{max}(S) + \mathrm{min}(S)},
\end{eqnarray}
is zero in this case.

The equivalent quantum circuit for the closed interferometer is shown in Fig.
\ref{fig1}f.  After the second Hadamard one obtains the state,
$\vert \psi_\mathrm{w} \rangle = \cos\frac{\phi}{2} \vert 0 \rangle - i\sin\frac{\phi}{2} \vert 1 \rangle,$
up to a global phase. The intensities recorded by the two detectors are now,
\begin{eqnarray}
S_{\mathrm{w},0} & = & \langle \psi_\mathrm{w} \vert D_0 \vert \psi_\mathrm{w} \rangle = \cos^2\frac{\phi}{2} \;\; \mathrm{and} \nonumber \\
S_{\mathrm{w},1} & = & \langle \psi_\mathrm{w} \vert D_1 \vert \psi_\mathrm{w} \rangle = \sin^2\frac{\phi}{2}.
\label{sw0}
\end{eqnarray}
Thus as a function of $\phi$, each detector obtains an interference pattern 
with visibility $\nu = 1$.  This setup clearly demonstrates the wave nature of the target qubit.

In the circuit corresponding to 
the Wheeler's experiment (Fig. \ref{fig1}g), the decision to insert or not to
insert the second Hadamard gate is to be made after the first Hadamard gate has been applied.

In this article we focus on the next modification, that is the quantum delayed-choice
experiment \cite{ionicioiu}.  
In the equivalent quantum circuit (Fig. \ref{fig1}h), the second Hadamard gate is to be
decided in a quantum way.  This involves an ancilla spin prepared in a
superposition state $\cos \alpha \vert 0 \rangle + \sin \alpha \vert 1 \rangle $.
This state can be prepared by rotating the initial $\vert 0 \rangle$ state of ancilla 
by an angle $\alpha$ about $y$-axis (using operator $Y_\alpha = e^{-i \alpha \sigma_y}$).
The second Hadamard gate is set to be controlled by the ancilla qubit.  If the ancilla is in state
$\vert 0 \rangle$, no Hadamard gate is applied, else if the ancilla is in state 
$\vert 1 \rangle$, Hadamard gate is applied.  The combined state of the two-qubit system
after the control-Hadamard gate is
\begin{eqnarray}
\vert \psi_\mathrm{wp,\alpha} \rangle = \cos \alpha \vert \psi_\mathrm{p} \rangle \vert 0 \rangle 
+ \sin \alpha \vert \psi_\mathrm{w} \rangle \vert 1 \rangle,
\end{eqnarray}
wherein the second ket denotes the state of ancilla.
After tracing out the ancilla, the reduced density operator for the system becomes,
\begin{eqnarray}
\rho_\mathrm{wp} = \cos^2 \alpha \vert \psi_\mathrm{p} \rangle \langle \psi_\mathrm{p} \vert + 
\sin^2 \alpha \vert \psi_\mathrm{w} \rangle \langle \psi_\mathrm{w} \vert.
\label{psiwp}
\end{eqnarray}
Again, the intensity recorded by each detector can be obtained by calculating the expectation values.
For example, the intensity at the detector D0 is,
\begin{eqnarray}
S_\mathrm{wp,0} (\alpha,\phi) &=& \mathrm{tr}[D_0 \; \rho_\mathrm{wp}] \nonumber \\
 &=&  \mathrm{tr}[D_0 \vert \psi_\mathrm{p} \rangle \langle \psi_\mathrm{p} \vert] \cos^2 \alpha
 + \nonumber \\
 && \mathrm{tr}[D_0 \vert \psi_\mathrm{w} \rangle \langle \psi_\mathrm{w} \vert] \sin^2 \alpha \nonumber \\
  & = &  S_{\mathrm{p},0} \cos^2 \alpha +  S_{\mathrm{w},0} \sin^2 \alpha \nonumber \\
  & = & \frac{1}{2} \cos^2 \alpha + \cos^2\frac{\phi}{2} \; \sin^2 \alpha.
\label{swp0}
\end{eqnarray}
It can be immediately seen that the visibility $\nu$ for the above interference varies as $\sin^2 \alpha$.
When $\alpha = 0$, the quantum system has a particle nature and when $\alpha = \pi/2 $ it
has a wave nature.  In the intermediate values of $\alpha$, the quantum system is morphed 
in between the particle and the wave nature.  
In the following section we describe the experimental demonstration of morphing of a quantum system between wave and particle
behaviors.

\section{Experiment}
\begin{center}
\begin{figure}
\includegraphics[width=7cm]{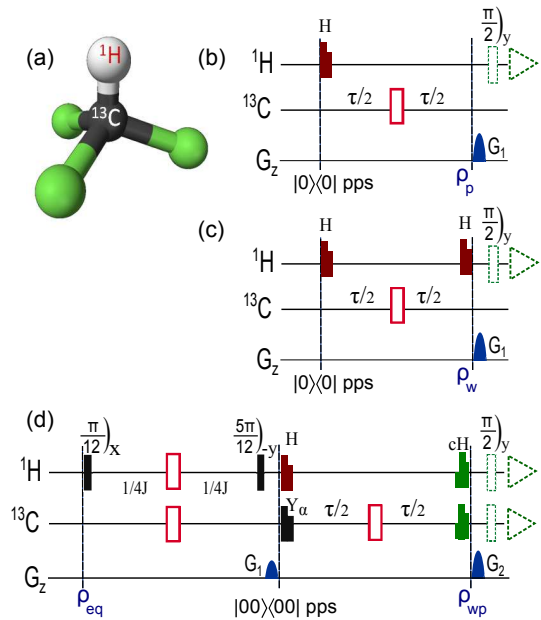}
\caption{Molecular structure of chloroform (a) and pulse-sequences (b-d)
for different setups of MZI.
Figs. (b) and (c) correspond to the open and closed setups respectively,
and (d) corresponds to the quantum delayed-choice experiment.
The unfilled rectangles are $\pi$ pulses.  Shaped pulses are strongly modulated pulses corresponding
to Hadamard gate (H), Y$_\alpha$ gate, and control-Hadamard (cH) gate. 
$\pi/2$ detection pulses are shown in dotted rectangles.
$J$ is the coupling constant and $\tau$ is the phase-shifting delay.
G$_1$ and G$_2$ are two pulsed-field-gradients for destroying coherences.
In (d) two separate experiments for $^1$H and $^{13}$C are recorded
after applying respective $\pi/2$ detection pulses.  
$\rho_\mathrm{eq}$, 
$\rho_\mathrm{p} = \vert \psi_\mathrm{p} \rangle \langle \psi_\mathrm{p} \vert$, 
$\rho_\mathrm{w} = \vert \psi_\mathrm{w} \rangle \langle \psi_\mathrm{w} \vert$, 
and 
$\rho_\mathrm{wp} = \vert \psi_\mathrm{wp} \rangle \langle \psi_\mathrm{wp} \vert$ represent
the states at different time instants. 
}
\label{fig2}
\end{figure}
\end{center}

The sample consisted of $^{13}$CHCl$_3$ (Fig. \ref{fig2}a) 
dissolved in CDCl$_3$.    
Here $^1$H and $^{13}$C spins are used as the target and the ancilla qubits respectively.
The two spins are coupled by indirect spin-spin interaction with a coupling constant of $J=209$ Hz.
All the experiments
were carried out at an ambient temperature of 300 K in a 500 MHz Bruker NMR spectrometer.

\subsection{Open and closed interferometers}
The pulse-sequences corresponding to open and closed setups of MZI
are shown in Fig. \ref{fig2}(b-c).  In these cases, the circuits (Fig. 1(e-f)) need
only a single target qubit and no ancilla qubit. Here $^1$H spin
is used as the target qubit, and its interaction with $^{13}$C spin is refocused 
during the MZI experiments.
Ideally both of these
setups need initializing the target qubit to $\vert 0 \rangle$ state.
In thermal equilibrium at temperature $T$ and magnetic field $B_0$, an ensemble of 
isolated spin-1/2 nuclei 
 exists in a Boltzman mixture,
\begin{eqnarray}
\rho_\mathrm{eq} =\frac{1}{2} e^{\epsilon/2} \vert 0 \rangle \langle 0 \vert
+ \frac{1}{2} e^{-\epsilon/2} \vert 1 \rangle \langle 1 \vert,
\end{eqnarray}
$\epsilon = \gamma \hbar B_0/kT$ is a dimensionless constant 
which depends on the magnetogyric ratio $\gamma$ of the spin.  
At ordinary NMR conditions $\epsilon \sim 10^{-5}$
and therefore $\rho_\mathrm{eq}$ is a highly mixed state.  Since preparing a pure 
$\vert 0 \rangle$ state requires extreme conditions, one can alleviate this problem
by rewriting the equilibrium state as the pseudopure state
\begin{eqnarray}
\rho_\mathrm{eq} =
\vert 0 \rangle \langle 0 \vert_\mathrm{pps} \approx
\frac{1}{2}\left(1-\frac{\epsilon}{2} \right)\mathbbm{1} + 
\frac{\epsilon}{2}\vert 0 \rangle \langle 0 \vert.
\end{eqnarray}
The identity part does neither evolve under the Hamiltonians, nor does it give raise
to NMR signals, and is therefore ignored.  
Thus the single qubit equilibrium state effectively mimics the state $\vert 0 \rangle$.

In all the cases (Fig. 2(b-d)), the first Hadamard gate on the target qubit is followed
by the phase shift. 
A 100 Hz resonance off-set of $^1$H spin was used to introduce the desired phase shift $\phi(\tau) = 200 \pi \tau$,
with the net free-precession delay $\tau$.  Experiments were carried out at 21 linearly spaced 
values of $\phi$ in the range $[0,2\pi]$.
The  $^{13}$C spin was set on-resonance and the 
$J$-evolution during $\tau$ was refocused with a $\pi$ pulse on $^{13}$C. 

\begin{center}
\begin{figure}[bottom]
\includegraphics[width=5cm,angle=-90]{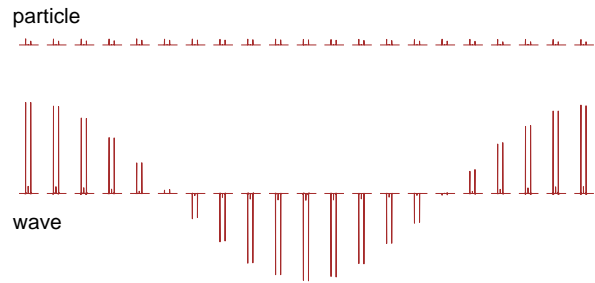}
\caption{The experimental spectra obtained after the open (top trace) and 
closed (bottom trace) setups of MZI.  
Each spectrum (pair of lines) corresponds to one of the 21 linearly spaced 
values of $\phi$ in the range $[0,2\pi]$.
}
\label{fig3}
\end{figure}
\end{center}

Unlike the open interferometer (Fig. 2b), the closed interferometer (Fig. 2c) has a second Hadamard 
gate.  In both of these cases, the intensity recorded by D1 detector corresponds to
the expectation value of $D_0 = \vert 0 \rangle \langle 0 \vert$ operator, which is a 
diagonal element of the density operator.  To measure this element, we destroy all the
off-diagonal elements (coherences) using a pulsed field gradient (PFG) G$_1$, followed by a $(\pi/2)_y$
detection pulse.  The most general diagonal density operator for a single qubit 
is
$\rho = \frac{1}{2} \mathbbm{1} + c \sigma_z$,
where $c$ is the unknown constant to be determined. After applying the $(\pi/2)_y$
detection pulse, we obtain $\frac{1}{2} \mathbbm{1} + c \sigma_x$.  The
corresponding NMR signal is proportional to $c$. The experimental NMR
spectra for the open and closed setups are shown in Fig. 
\ref{fig3}.  These spectra are normalized w.r.t. equilibrium detection.
Since both the pathways created by BS1 are equally probable in the open
MZI, $c=0$ and therefore spectrum vanishes.  On the other hand, 
because of the second beam-splitter (BS2) in closed MZI, $c$ becomes 
$\phi$ dependent, and hence the interference pattern.

\begin{center}
\begin{figure}
\includegraphics[width=6cm]{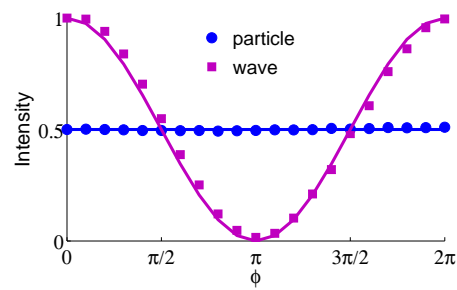}
\caption{The experimental intensities $S_\mathrm{p,0}$ (particle) and 
$S_\mathrm{w,0}$ (wave) at various values of $\phi$.
}
\label{fig4}
\end{figure}
\end{center}

The corresponding intensities $S_\mathrm{p(w),0} = c+1/2$ are shown in
Fig. \ref{fig4}.  
The theoretical values from expressions (\ref{sp0}) and (\ref{sw0}) are also shown
in solid lines. 
The experimental visibility of interference in the 
particle case is 0.02 and that in the wave case is 0.97.
As explained in the previous section, the open setup demonstrates the particle nature
and the closed setup demonstrates the wave nature.

\subsection{Quantum delayed-choice experiment}
The circuit for quantum delayed-choice experiment is shown in Fig. \ref{fig1}h and
the corresponding NMR pulse-sequence is shown in Fig. \ref{fig2}d.  This circuit
requires one target qubit ($^1$H) and one ancilla qubit ($^{13}$C). 
The equilibrium state of the two-qubit system does not
correspond to a pseudopure state and therefore it is necessary to redistribute the 
populations to achieve the desired pseudopure state.
We used spatial averaging technique to prepare the pseudopure state \cite{cory}
\begin{eqnarray}
\rho_\mathrm{pps} = \frac{1-\epsilon'}{4}\mathbbm{1} + \epsilon' \vert 00 \rangle \langle 00 \vert,
\end{eqnarray}
where $\epsilon'$ is the residual purity.

All the gates on the target and the ancilla were realized using
strongly modulated pulses (SMPs) \cite{fortunato,maheshsmp}.
The SMPs were constructed to be robust against radio-frequency inhomogeneities
with an average Hilbert-Schmidt fidelity of over 0.995 
over $\pm10\%$ distribution of radio-frequency amplitudes.  
After the control-Hadamard gate, the state of the two-qubit system is expressed by
the density operator $\rho_\mathrm{wp}$ (eqn. \ref{psiwp}) up to the unit background.

\begin{center}
\begin{figure}
\includegraphics[width=6.1cm,angle=-90]{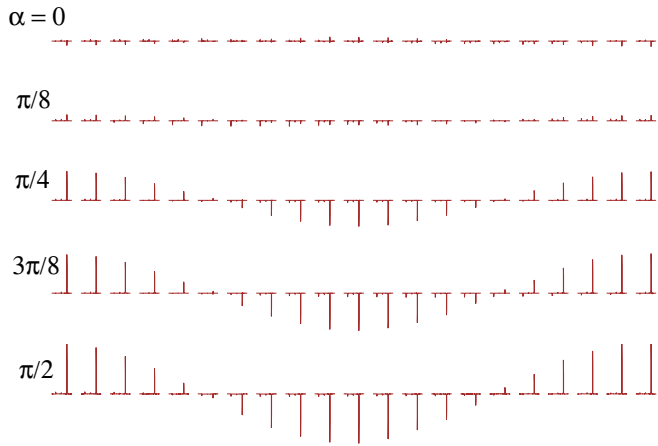}
\caption{
The experimental spectra obtained after the quantum delayed choice
experiment with $(\pi/2)_y$ detection pulse on target ($^1$H) qubit.
These spectra are recorded with 21 equally spaced values of $\phi \in [0,2\pi]$ and
at different $\alpha$ values (as indicated).  In each spectrum, only one
line is expected due to the preparation of pseduopure state.
}
\label{fig5}
\end{figure}
\end{center}

The interference $S_\mathrm{wp,0}$ (in eqn. \ref{swp0}) due to the detection operator 
$D_0 = \vert 00 \rangle \langle 00 \vert$ 
can be obtained by measuring the first diagonal element of the density matrix, and
hence complete density matrix tomography is not necessary \cite{soumyajmr}.
As in the single qubit case, we apply a PFG $G_2$ which averages out all the coherences and 
retains only the diagonal part of the density matrix.  The most general diagonal
density matrix of a two-qubit system is of the form
\begin{eqnarray}
\rho = \frac{1}{4}\mathbbm{1} \otimes \mathbbm{1} 
+ c_1 \sigma_z \otimes  \mathbbm{1}  
+ c_2 \mathbbm{1} \otimes  \sigma_z  
+ c_3 \sigma_z \otimes  \sigma_z,
\end{eqnarray}
with the unknown constants $c_1$, $c_2$, and $c_3$.

Recording the target spectrum after a $(\pi/2)_y$ pulse on the above state
gives two signals proportional to $c_1+c_3$ and $c_1-c_3$.  
The spectra of the target qubit at various values of $\phi$ and
$\alpha$ are shown in Fig. \ref{fig5}.  
The signals obtained after applying a $(\pi/2)_y$ pulse on either qubit after
preparing the $\vert 00 \rangle$ pseudopure state are used to normalize these
intensities.  In each spectrum, the left transition (corresponding to the $\vert 0 \rangle$
state of ancilla), vanishes because of the particle nature (similar to the top trace of Fig. \ref{fig3})
and the right transition (corresponding to the $\vert 1 \rangle$ state of ancilla) displays the 
interference pattern because of the wave nature (similar to the bottom trace of Fig. \ref{fig3}).

\begin{center}
\begin{figure}
\includegraphics[width = 5.5cm]{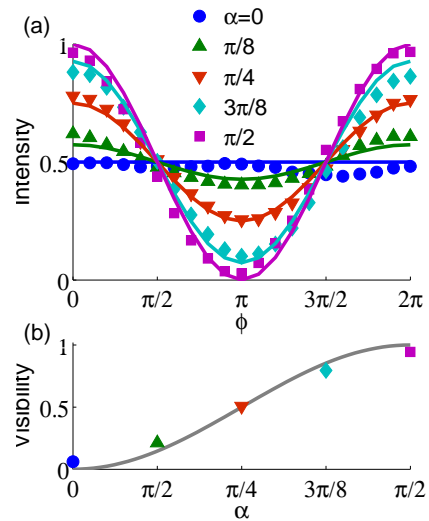}
\caption{The intensities $S_\mathrm{wp,0}(\alpha,\phi)$ versus phase $\phi$ for
different values of $\alpha$ (a) and the visibility $\nu$ versus $\alpha$ (b).
The theoretical values are shown in solid lines and the experimental results
are shown by symbols.  
}
\label{fig6}
\end{figure}
\end{center}

Similarly, recording the ancilla spectrum after a $(\pi/2)_y$ pulse 
gives two signals proportional to $c_2+c_3$ and $c_2-c_3$.  
From these four transitions one can precisely determine all the three unknowns $c_1$,
$c_2$, and $c_3$, and obtain the 
population $S_\mathrm{wp,0} = 1/4 + c_1+c_2+c_3$.
Calculated experimental intensities $S_\mathrm{wp,0}$ are shown in 
Fig. \ref{fig6}a. The intensities were measured for five
values of $\alpha$ in the range $[0,\pi/2]$, and for 21 values of $\phi$ in the range
$[0,2\pi]$.  The theoretical values from expression (\ref{swp0}) are also shown
in solid lines. 
The experimental values were found to have small random errors with a standard deviation less 
than 0.01.  
The significant systematic errors are due to experimental limitations
such as radio-frequency inhomogeneity and spectrometer non-linearities.

The visibility $\nu$ calculated at different values of $\alpha$ are plotted in 
Fig. \ref{fig6}b.  The theoretical visibility varies as $\sin^2\alpha$ as
explained in the section II.
There appears a general agreement between the
quantum mechanical predication (solid-line) and the experiments (symbols).

\section{Conclusions}
We have studied the open and closed setups of Mach-Zehnder interferometer using
nuclear spin qubits, and demonstrated the particle-like and wave-like behaviors
of the target qubit. 
Previously NMR interferometer has been used to study dipolar oscillations in solid state NMR \cite{stoll}
and to measure geometric phases in multi-level systems \cite{dietergp,arindamgp,dugp}. 
We have reported the first experimental demonstration of the quantum delayed-choice
experiment using NMR interferometry.  

Bohr's complementarity principle is based on mutually exclusive experimental arrangements.
However, the quantum delayed-choice experiment proposed by Ionicioiu and Terno 
\cite{ionicioiu}, 
suggests that we can study the complementary properties like particle and wave behavior
of a quantum system in a single experimental setup if the ancilla is prepared
in a quantum superposition.  This experiment is the quantum version of the 
Wheeler's delayed-choice experiment.
The quantum delayed-choice experiment suggests a reinterpretation of complementarity principle:
instead of complementary experimental setups, the new proposal suggests 
complementarity in the experimental data.  

The NMR systems provide perfect platforms for studying such phenomena.
In our experiments we found a general agreement between the intensities and
the visibilities of the interference with the theoretically expected values.
These experiments not-only confirm the intrinsic wave-particle duality of
quantum systems, but also demonstrates continuous morphing of quantum systems
between wave and particle behavior of the target qubit depending on the quantum state 
of the ancilla qubit.

\section*{Acknowledgment}
Authors gratefully acknowledge Prof. Anil Kumar and
Dr. Vikram Athalye for discussions.  This work was supported 
by DST project SR/S2/LOP-0017/2009. 
S.S.Roy acknowledges CSIR for research fellowship.

\end{document}